\documentclass{elsart}
\usepackage{amssymb}
\usepackage{amsmath}
\usepackage{graphics}
\usepackage{epsfig}
\usepackage{amssymb}
\input epsf

\newcommand{\dd}{\mbox{\rm d}}
\newcommand{\pppi}{\mbox{$pp\to \{pp\}_{\!s\,}\pi^0$}}
\newcommand{\pnpi}{\mbox{$pp\to \{pn\}_{\!s\,}\pi^+$}}
\newcommand{\dpi}{\mbox{$pp\to d\pi^+$}}
\begin{document}
\begin{frontmatter}
\title{Production of the $\mathbf{^1S_0}$ diproton in
the $\mathbf{pp\rightarrow pp\pi^0}$ reaction at 0.8~GeV}
\vspace{-5mm}

\author[dubna]{S.~Dymov},
\author[dubna]{D.~Gusev},
\author[ikp]{M.~Hartmann},
\author[ikp]{V.~Hejny},
\author[erlangen,tbilisi]{A.~Kacharava},
\author[munster]{A.~Khoukaz},
\author[dubna]{V.~Komarov},
\author[cracow]{P.~Kulessa},
\author[dubna]{A.~Kulikov},
\author[dubna]{V.~Kurbatov},
\author[munster]{N.~Lang},
\author[dubna,tbilisi]{G.~Macharashvili},
\author[munster]{T.~Mersmann},
\author[dubna,ikp]{S.~Merzliakov},
\author[gatchina]{S.~Mikirtytchiants},
\author[ikp]{A.~Mussgiller},
\author[ikp]{D.~Prasuhn},
\author[ikp]{F.~Rathmann},
\author[ikp]{R.~Schleichert},
\author[ikp]{H.~Str\"oher},
\author[dubna]{Yu.~Uzikov},
\author[london]{C.~Wilkin}, and
\author[erlangen,dubna]{S.~Yaschenko}.

\address[dubna]{Laboratory of Nuclear Problems, Joint Institute for
Nuclear Research, 141980 Dubna, Russia}
\address[ikp]{Institut f\"ur Kernphysik, Forschungszentrum J\"ulich, 52425 J\"ulich, Germany}
\address[erlangen]{Physikalisches Institut II, Universit\"at
Erlangen--N\"urnberg, 91058 Erlangen, Germany}
\address[tbilisi]{High Energy Physics Institute, Tbilisi State
  University, 0186 Tbilisi, Georgia}
\address[munster]{Institut f\"ur Kernphysik, Universit\"at M\"unster,
48149 M\"unster, Germany}
\address[cracow]{Institute of Nuclear Physics, 31342 Cracow, Poland}
\address[gatchina]{St. Petersburg Nuclear Physics Institute, 188350 Gatchina, Russia}
\address[london]{Physics and Astronomy Department, UCL, London, WC1E 6BT, UK}


\begin{abstract}
The $pp\to pp\pi^0$ differential cross section has been measured
with the ANKE spectrometer at COSY--J\"ulich for pion cms angles
between $0^\circ$ and $15.4^\circ$ at a proton beam energy of
0.8~GeV. The selection of diproton pairs with an excitation energy
$E_{pp}<3$~MeV ensures that the final $pp$ system is dominantly in
the spin--singlet $^1\!S_0$ state. The kinematics are therefore
very similar to those of $pp\to d\pi^+$ but with different spin
and isospin transitions. The results will thus provide a crucial
extra test of pion production models in nucleon--nucleon
collisions.

The cross sections, which are over two orders of magnitude smaller
than those of \dpi, show a forward dip, even stronger than that
seen at lower energies. This behaviour is well reproduced in a
theoretical model that includes $P$--wave $\Delta N$ states.
\end{abstract}

\end{frontmatter}
%
%

Single pion production in nucleon--nucleon collisions, $NN\to
NN\pi$, is the first inelastic process that can be used to test
our understanding of the underlying meson--baryon dynamics of the
$NN$ interaction~\cite{Garcilazo,machnerjh,HanhartPR}. By far the
cleanest reaction to study is \dpi, where the differential cross
section and multitude of spin observables that have been measured
over the years~\cite{Arndt} can confront the different theoretical
models.

In contrast, very little was known about the $pp\to pp\pi^0$
reaction, though the unexpectedly large $\pi^0$ production rate
observed near threshold~\cite{meyer92} led to a flurry of
theoretical activity~\cite{Riska}. Now in cases where the
excitation energy $E_{pp}$ of the final protons is very small, due
to the Pauli principle, this reaction will excite only the
$J^p=0^+$ ($^1\!S_0$) diproton state. Despite having kinematics
very similar to those of \dpi, this reaction involves different
transitions in the $NN$ system and, in particular, the role of the
$\Delta$ isobar is expected to be much suppressed because the
$S$--wave $\Delta N$ intermediate state is forbidden.

Some information on the transitions involved has been extracted
from quasi--free pion absorption on $^3$He~\cite{Moinester}.
However, previous measurements of the differential cross sections
for \pppi\ have only been carried out up to a beam energy of
425$\:$MeV, with a cut imposed on the excitation energy of
$E_{pp}<3\:$MeV~\cite{maeda,bilger}. This value satisfies the
requirement that the spin--singlet $S$-wave ($^1\!S_0$) final
state, here denoted by $\{pp\}_{\!s}$, should dominate while
providing reasonable statistics. One important feature of the
experimental data is that with the $E_{pp}$ selection the cross
sections show a forward dip whereas, if no cut is applied on the
excitation energy, then for the higher beam energies there is a
forward maximum~\cite{bilger,rappenecker}.

Now 425$\:$MeV is below the threshold of $N\Delta$ production; the
data show no sign of being influenced in any clear way by the
$\Delta$ and we need to go to higher energy to investigate the
effects of $P$--wave $\Delta N$ systems. As part of a programme to
investigate the small $E_{pp}$ region in intermediate energy
nuclear reactions, in particular in large momentum transfer
deuteron breakup reactions~\cite{komarov03,aypaper}, we have
carried out a high statistics measurement of the \pppi\ reaction
at $T_p=800\:$MeV for pion cm angles below $15.4^{\circ}$.

The experiment was performed at the magnetic spectrometer
ANKE~\cite{ref04}, placed at an internal target position of the
COSY COoler SYnchrotron~\cite{cosy}. Fast charged particles,
resulting from the interaction of the proton beam with the
hydrogen cluster--jet target~\cite{Munster}, were registered in
the Forward Detector (FD) system~\cite{Dymov}. Its hodoscope
provided a trigger signal and an energy--loss measurement. It also
allowed a determination of the differences in arrival times for
particle pairs hitting different counter elements. The tracking
system gave a momentum resolution $\sigma_p/p\approx 0.8$ --
$1.2\%$ for protons in the range (0.5 -- 1.2)$\:$GeV/c.
%
%
\begin{figure}[Hbt]
\begin{center}
\epsfig{figure=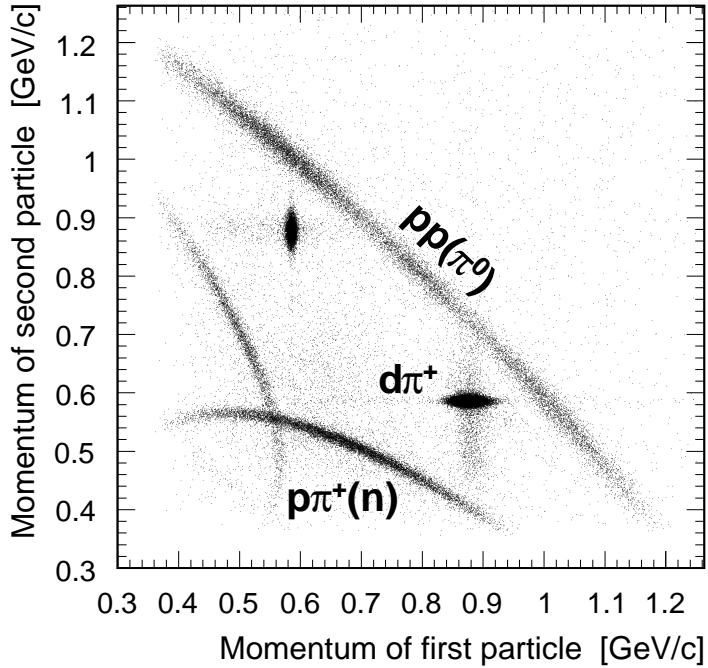,width=10cm} \caption{Scatter plot of the
magnitudes of the momenta of two charged particles detected in the
FD. The selection procedure introduces a slight bias as to which
particle is called ``1'', but this does not affect the subsequent
analysis.} \label{p1p2}
\end{center}
\end{figure}
%
%

The trigger used required the crossing of the two planes of the
scintillation hodoscopes by at least one charged particle but, in
the subsequent off--line analysis of the \pppi\ reaction, only
events with two tracks in the FD were retained. In Fig.~\ref{p1p2}
is shown a two--dimensional scatter plot of the magnitudes of
their two momenta corresponding to about half of our statistics.
Due to the limited angular acceptance of ANKE, there are kinematic
correlations for reactions with two and three particle in the
final state. One therefore sees in the figure islands
corresponding to \dpi\ and bands resulting from $pp\to pn\pi^+$
and $pp\to pp\pi^0$. Candidates for the latter reaction are well
separated from the other processes. Furthermore, in approximately
75\% of cases the particles hit different counters in the
hodoscope and the difference in their arrival time could also be
used in the selection. The $d\pi^+$ pairs coming from the $pp
\rightarrow d \pi^+$ reaction, which could potentially provide the
most serious physical background, are separated from the $pp$
pairs from $pp \rightarrow pp \pi^0$ in time difference by more
than at 8$\:$ns, whereas the actual resolution is better than
0.5$\:$ns

\begin{figure}[Hbt]
\begin{center}
\epsfig{figure=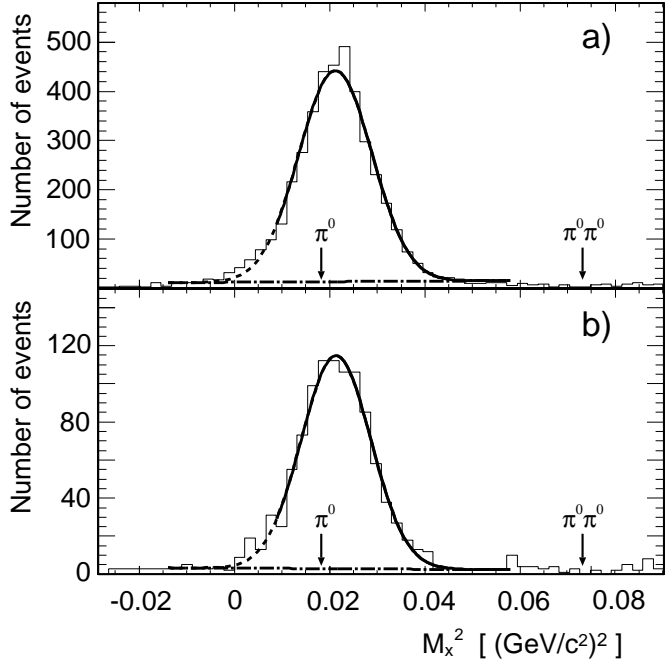,width=8.8cm} \caption{Distributions in the
square of the missing mass for candidates for the $pp\to ppX$
reaction with excitation energy $E_{pp} < 3\:$MeV and
$\theta_{pp}^{cm} \le 15.4^{\circ}$ when the protons (a) hit
different counters, and (b) the same counter. From the indicated
positions of the $\pi^0$ peak and the $2\pi^0$ threshold it is
seen that single and double pion production can be clearly
separated. Gaussian fits to the $\pi^0$ peak plus a constant
background yield a total number of $\pi^0$ events in (a) and (b)
of respectively 4425 and 1008.} \label{MM1}
\end{center}
\end{figure}

The distributions of missing mass squared, $M_X^2$, are shown
separately in Fig.~\ref{MM1} for single--counter and
double--counter candidates with low excitation energy in the $pp$
system, $E_{pp} < 3\:$MeV. In both cases one sees a very clean
$\pi^0$ peak centred at $0.021\:$(GeV/c$^2$)$^2$, which agrees
with $m_{\pi^0}^2$ to well within our experimental precision. The
widths of the Gaussian fits are compatible with those obtained
from Monte Carlo simulations; the marginally narrower peak in the
single--counter data is due to these events generally having a
smaller opening angle resulting in the kinematics being slightly
better defined. The backgrounds are small and  slowly varying and
two--pion production can be clearly excluded in either case. There
is a small excess of events observed on the left side of the
$\pi^0$ peak. These may correspond to single photon production
through $pp\to \{pp\}_{\!s\,}\gamma$ and so the regions indicated
by dashed lines in  Fig.~\ref{MM1} have not been included in the
Gaussian fits. Since this interpretation is not unambiguous and
these events might still correspond to good $\pi^0$ events, we
have added an extra 2\% to the systematic error. Given that the
two data sets are completely compatible, they have been grouped
together in the subsequent analysis. The resolution in excitation
energy for the combined $pp\to pp\pi^0$ events was $\sigma
(E_{pp})\approx 0.2$--$0.3\:$MeV for $E_{pp}$ in the range
0--3~MeV.

%
%
%
%

The value of the luminosity needed to determine the cross section
was found by comparing the yield of $pp$ elastic scattering,
measured simultaneously with the other reactions, with that
deduced from the SAID data base~\cite{ref06}. The integrated
luminosity obtained from this is $L_{\rm int} = (6.72 \pm 0.26)
\times 10^{34}\:$cm$^{-2}$, where the error comes mainly from
averaging over the angular bins.

%
%
\begin{figure}[Hbt]
\begin{center}
\epsfig{file=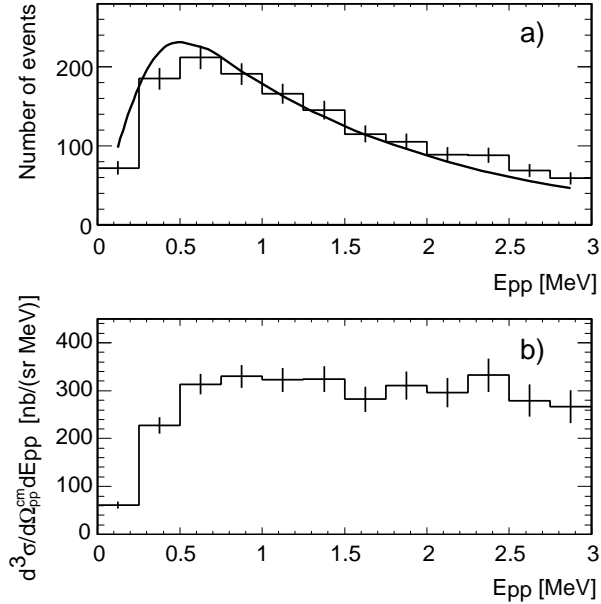,height=8cm}%
\caption{(a) Number of $pp\to pp\pi^0$ events in the interval
$8.9^{\circ}<\theta_{pp}^{cm}<13^{\circ}$ as a function of
$E_{pp}$; (b) The same data corrected for acceptance and detection
efficiency and presented as differential cross sections. Only
statistical errors are shown. The curve results from passing the
Migdal--Watson function $|T(E_{pp})|^2$ of Eq.~(\ref{fsi}),
multiplied by phase space, through a Monte Carlo simulation of the
ANKE apparatus and normalising the predictions to the summed
experimental histogram. Similar results are found for the other
angular intervals.\label{new1}}
\end{center}
\end{figure}
%
%

In order to determine the triply differential cross section
$\dd^3\sigma/(\dd\Omega_{pp}^{cm}\,\dd E_{pp})$, events selected
in the range $0\le E_{pp}< 3\:$MeV were divided into groups of
equal intervals in $\cos\theta_{pp}^{cm}$. The energy spectrum of
counts in the angular interval
$8.9^{\circ}<\theta_{pp}^{cm}<13^{\circ}$ is shown in
Fig.~\ref{new1}a. The spectra in the other intervals demonstrate a
similar behaviour. Values of the corresponding cross sections were
obtained from such a distribution by taking into account
geometrical acceptance, efficiency in the track recognition
algorithm for two particles, interactions in the constituent
materials, efficiency and resolution of the detectors and other
known effects on the basis of a Monte Carlo simulation of the ANKE
setup. This leads to the histogram with the statistical errors
presented in Fig.~\ref{new1}b.

The rapid rise of the spectrum with $E_{pp}$ from threshold
illustrated in Fig.~\ref{new1}b is typical of all intermediate
energy reactions where one produces proton pairs and is induced by
the $pp$ final state interaction. We have indeed observed exactly
the same phenomenon in the $pd\to (pp)n$ reaction with the same
apparatus at ANKE~\cite{komarov03,aypaper}. This effect is often
parameterised, in the Migdal--Watson
approximation~\cite{migdal-watson}, by the square of the low
energy $pp$ elastic scattering amplitude for which
\begin{equation}
\label{fsi}%
|T(E_{pp})|^2=\frac{1}{|C(\eta)|^2}\left(\frac{\sin\delta}{k}\right)^2,
\end{equation}
 where
$|C(\eta)|^2$ is the Coulomb penetration factor evaluated at
$\eta=\alpha m_p/2k=\alpha\sqrt{m_p/E_{pp}}/2$, and $\delta$ is
the combined Coulomb--nuclear phase shift. This has been evaluated
numerically for the Reid soft core potential, for which the
scattering length $a_{pp}=-7.8\:$fm.

\begin{figure}[Htb]
\begin{center}
\epsfig{figure=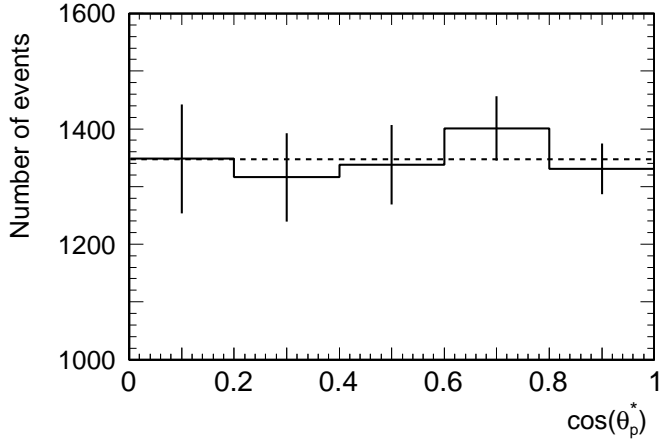,width=8.8cm}%
\caption { Distribution of acceptance--corrected \pppi\ events
with $E_{pp}<3\:$MeV over $\cos{\theta_{p}^{*}}$, where
$\theta_{p}^{*}$ is the angle between the directions of the proton
and diproton momenta in the centre of mass of the diproton.
Note that the vertical scale does not start from
zero.\label{c1dcos2a}}
\end{center}
\end{figure}

The Migdal--Watson factor of Eq.~(\ref{fsi}) was used as an event
generator together with phase space to
provide candidates which were then traced through the experimental
setup, taking into account all its known features. The resulting
smoothed curve, shown in Fig.~\ref{new1}a, provides a
semi--quantitative description of the data which is quite
sufficient for our purpose, where we quote cross sections summed
over energy. The data are a little above the curves at the higher
$E_{pp}$ and we cannot exclude some small $P$--wave contribution
though globally the angular distribution of the $pp$ system in its
rest frame shown in Fig.~\ref{c1dcos2a} is consistent with isotropy.
It should be noted that the $^{1\!}P_0$ final state would also produce
a flat distribution.

\begin{figure}[Htb]
\begin{center}
\epsfig{figure=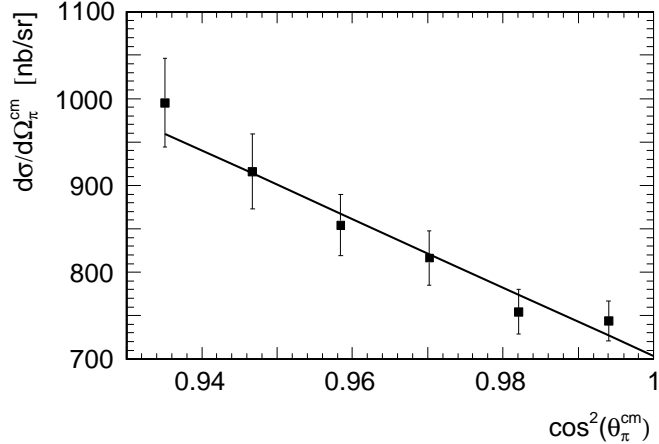,width=8.8cm}%
\caption {The measured \pppi\ differential cross section for
$E_{pp}<3\:$MeV as a function of $\cos^2(\theta_{\pi}^{cm})$. The
curve is a straight--line fit to the data.\label{c1dcos2b}}
\end{center}
\end{figure}

Due to the identity of the initial protons, the differential cross
section is an even function of $\cos\theta_{\pi}^{cm}$ and in
Fig.~\ref{c1dcos2b} it is plotted \emph{versus}
$\cos^2\theta_{\pi}^{cm}$. The results show a monotonic decrease
towards the forward direction and, as seen from the figure, they
can be well parameterised by the linear function
$a(1+b\sin^2\theta_{\pi}^{cm})$, where $a=(704\pm22_{\rm stat}\pm
32_{\rm syst})\:$nb/sr and $b=5.6\pm 1.2$. With the same $E_{pp}$
cut as used here, a similar forward dip was observed in this
reaction at lower energies, $T_p\le 425\:$MeV~\cite{maeda,bilger},
though for these energies it was found that $b$ was much smaller,
being always less than 1.4. Since, for such small values of
$E_{pp}$, the final diproton must be dominantly in an $S$--wave,
constraints from spin--parity and Fermi statistics then require
the pion to be in an even partial wave. As a consequence, the
forward dip was attributed to an interference between the pion $s$
and $d$ waves~\cite{bilger}. Given that the influence of
$d$--waves might be expected to increase with energy, it is
perhaps not surprising that we find a larger slope parameter at
800$\:$MeV.

Preliminary theoretical predictions have been made for the \pppi\
differential cross section at 800$\:$MeV in a model that includes
contributions from $P$--wave $\Delta N$ intermediate
states~\cite{Jouni,Jouni2}. The overall magnitude is similar to
that which we have observed and, in particular, the forward slope,
driven by the pion $d$--wave, is well reproduced. It is expected
that our data, combined with those at lower energies, will allow
such models to be refined.

Though we have argued that the kinematics of \pppi\ and $pp\to
d\,\pi^+$ are quite similar, the underlying dynamics must be very
different. This is illustrated in Table~\ref{tab}, where we show
the values of the two differential cross sections and their ratio
$R(\pi^0/\pi^+)$ in the forward direction obtained at different
energies. The results seem to indicate that there might be a broad
minimum in $R$ in the $\Delta$ region of the $pp\to d\pi^+$
reaction.

\begin{table}[ht]
\begin{center}
\begin{tabular}{|c|c|c|c|} \hline
$T_p$&$\sigma(pp\pi^0$)&$\sigma(d\pi^+$)&$R(\pi^0/\pi^+)$ \\
(MeV)&(nb/sr)&($\mu$b/sr)&$\times 10^3$ \\
\hline
 310 &  $109\pm8$           & $14.1\pm0.3$ & $7.7\pm0.6$  \\
 320 &  $110\pm6$           & $22.1\pm0.6$ & $5.0\pm0.3$  \\
 340 & $\phantom{1}95\pm10$ & $41.9\pm1.5$ & $2.3\pm0.3$  \\
 360 & $\phantom{1}86\pm7$  & $67.1\pm2.7$ & $1.3\pm0.1$  \\
 400 &  $121\pm9$           & $135\pm2$    & $0.9\pm0.1$  \\
 425 &  $171\pm14$          & $189\pm3$    & $0.9\pm0.1$  \\
 800 &  $704\pm22$          & $155\pm4$    & $4.5\pm0.4$  \\ \hline
\end{tabular}
\vspace{5mm}
\caption{Zero degree differential cross sections for
\pppi\ with $E_{pp}<3\:$MeV from the present experiment at
800$\:$MeV with the lower energy data being taken from Ref.~\cite{bilger}.
The values of the \dpi\ cross sections are obtained from the SAID
SP96 solution with the range of the other solutions being taken as a
rough estimate of the error bars~\cite{ref06}. The ratio
$R(\pi^0/\pi^+)$ of the two pion--production cross sections is
also presented. \label{tab}}
\end{center}
\end{table}

The GEM collaboration has recently published high precision data
on the ratio of the forward production of pions in the
$pp\to\pi^+d$ and $pp\to\pi^+pn$ reactions at
981$\:$MeV~\cite{GEM}. Since the spin--singlet $pn$ final state
interaction has a much sharper energy dependence than that of the
triplet, from the shape of the pion momentum spectrum they could
put an upper limit on the amount of $pn$ singlet produced.
Integrating over excitation energies $E_{pn}<3\:$MeV, it is seen
that the ratio of $\displaystyle
\frac{\dd\sigma}{\dd\Omega}(\pnpi)$ to $\displaystyle
\frac{\dd\sigma}{\dd\Omega}(\dpi)$ could be at most about 5\%. If
Coulomb effects are ignored, the cross sections for spin--singlet
production through \pnpi\ and \pppi\ should be identical. Though
Coulomb suppression in the $\{pp\}_{\!s}$ final state will be
large, it looks very doubtful whether the study of the spectrum
alone will be sufficient to isolate the \pnpi\ cross section in
view of the values presented in Table~\ref{tab}. Measuring the
proton and pion in coincidence, as has been done for example in
Refs.~\cite{HG,Abaev} and analysed in a model--independent
way~\cite{uzwilkin}, still only provides upper bounds. The study
of $\pi^0$ production therefore seems to be the most realistic way
of investigating the $^1\!S_0$ final state here.

Data on quasi--free pion production in the $pd\to\{pp\}_{\!s\,}X$ reaction
were obtained as a by--product of our deuteron break-up
measurements~\cite{komarov03} and these results will be interpreted
in terms of the sum of the cross sections for \pppi\ and
$pn\to \{pp\}_{\!s\,}\pi^-$. At 800$\:$MeV it will then be possible to
subtract the $\pi^0$ contribution reported here in order to obtain
data on $\pi^-$ production.

It is intriguing to note that a very similar ratio to that of
Table~\ref{tab} has been observed for backward dinucleon
production in the $pd\to \{pp\}_{\!s\,}n$ and $pd\to dp$ reactions
at intermediate energies~\cite{komarov03}. Now such a connection
would be natural within a one--pion--exchange mechanism, where the
large momentum transfer $pd\to dp$ reaction is driven by a \dpi\
sub-process~\cite{Craigiewilkin,Kolybasovsm}. More quantitative
estimates of the $pd\to \{pp\}_{\!s\,}n$ cross section, where the
\pppi\ sub-process is used rather than the \dpi, are currently
under way~\cite{YuriColin}.

It is seen from Table~\ref{tab} that there is a real lack of data
on the \pppi\ reaction in the $\Delta$ region and this could be
usefully filled by further experiments at ANKE. It should also be
noted that, unlike the complicated spin structure connected with
the \dpi\ reaction, only two spin amplitudes which are functions
of $\cos^2\theta_{\pi}$ are required to describe the \pppi\
reaction. These can be isolated, up to an unmeasurable overall
phase, by determining the proton analysing power and the initial
$pp$ spin correlation $C_{xx}$. Both of these experiments can be
carried out at small angles using ANKE~\cite{SPIN} and the
resulting amplitude analysis will tie
down even further $\pi NN$ dynamics at intermediate energies.\\

This work was supported in part by the BMBF grants ANKE
COSY--JINR, Kaz-02/001 and Heisenberg--Landau programme. We are
grateful to many other members of the ANKE collaboration who
provided strong support for the measurement. Important discussions
with J.A.~Niskanen and correspondence with I.~Strakovsky are also
gratefully acknowledged.

\end{document}